\documentclass[a4paper,12pt]{article}
\usepackage{amsmath,amssymb,amsfonts}
\usepackage{feynmf}
\usepackage[pdftex]{graphicx,color}
\usepackage[toc,page]{appendix}
\usepackage{flafter}
\usepackage{hyperref}
\usepackage{subfig}
\usepackage[section]{placeins}
\usepackage{slashed}
\usepackage{fullpage}
\numberwithin{equation}{section}
\usepackage{enumerate}

\makeatletter

\newcommand{\Rmnum}[1]{\expandafter\@slowromancap\romannumeral #1@}
\newcommand\email[1]{{\tt\href{mailto:#1}{#1}}}

\makeatother

\begin{document}

\begin{titlepage}

\setcounter{page}{0}

\vspace*{-2cm}

\begin{flushright}
{Edinburgh 2011/19}
\end{flushright}

\vspace{0.6cm}

\begin{center}
{\Large \bf Lattice Determination of the Hadronic Contribution to the Muon $g-2$
using Dynamical Domain Wall Fermions} \\

\vskip 0.8cm

{\bf Peter Boyle, Luigi Del Debbio, Eoin Kerrane, \& James Zanotti}\\
{\sl
School of Physics and Astronomy,\\
University of Edinburgh,\\
Edinburgh EH9 3JZ,\\
Scotland\\
E-mail: \email{paboyle@ph.ed.ac.uk}, \email{luigi.del.debbio@ed.ac.uk}, \email{eoin.kerrane@ed.ac.uk}, \email{jzanotti@ph.ed.ac.uk}
}

\vskip .6cm

\end{center}

\begin{abstract}
We present a calculation of the leading order hadronic contribution to the
anomalous magnetic moment of the muon for a dynamical simulation of 2+1 flavour
QCD using domain wall fermions. The electromagnetic 2-point function is
evaluated on the RBC-UKQCD lattice gauge configurations and this is fitted to a continuous
form motivated by models of vector dominance. We determine a robust and reliable technique for performing this fit, allowing us to extract the most accurate
results possible from our ensembles. This combined with data at very light quark masses produces the result
\begin{equation*}
 a_\mu^{(2)had}=641(33)(32)\times 10^{-10}
\end{equation*}
at the physical point, where the first uncertainty is statistical, and the second is an estimate of systematics, which is in
agreement with previous results. We outline various methods by which this calculation can and will be improved in order to compete with the accuracy of
alternative techniques of deducing this quantity from experimental scattering data.
\end{abstract}

\vfill

\end{titlepage}

\section{Introduction}

The anomalous magnetic moment $a$ of a lepton, is half the discrepancy from 2
($a=\frac{g-2}{2}$) of $g$, the gyromagnetic ratio or Land\'{e} $g$-factor,
which relates the spin $\vec{S}$ of the lepton to its magnetic moment
$\vec{\mu}$ as
\begin{equation}
 \vec{\mu}=g\frac{e}{2m}\vec{S}.
\end{equation}
It is given the name ``anomalous'' because it is a purely quantum effect and so
is zero in a classical theory. 

\sloppy The one-loop computation of the electron anomalous magnetic moment $a_e$ by
Schwinger \nolinebreak\cite{Schwinger:1948iu} was one of the first such calculations, and
provided strong evidence in support of the young theory of quantum
electrodynamics (QED) by explaining observed hyper-fine phenomena which
were not well understood. Since then $a_e$ has become possibly the most
accurately determined quantity in science, being known to a precision better
than one part per billion \cite{Hanneke:2008tm}. The corresponding theoretical
calculation has achieved similar accuracy \cite{Laporta:2008zz}. Because of the
relatively light mass of the electron, the calculation is strongly dominated by
QED contributions with virtual electrons, which are known to a good accuracy to
four-loops. Using an independent determination of the fine-structure constant
$\alpha$ from atomic interferometry results in a value of $a_e$ which agrees
with the experimental result, with an uncertainty over 30 times greater.
Combining the experimental and theoretical results for $a_e$
in terms of the fine structure constant $\alpha$ provides the most accurate available determination of
$\alpha$ \cite{Hanneke:2008tm}.

Because of its heavier mass, $\frac{m_\mu^2}{m_e^2}\simeq40000$, the muon
anomalous magnetic moment $a_\mu$ is far more sensitive to contributions from
other sectors of the standard model, as well as to any potential new-physics contributions.
This makes it a far more robust test of the standard model, and a much more
interesting  searching-ground for signals of new physics. The current
experimental result, while not nearly as accurate as that for $a_e$ is still
remarkably precise \cite{Bennett:2006fi}:
\begin{equation}
a_\mu=11 659 208.0(6.3) \times 10^{-10},
\end{equation}
which remains a precision of better than one part per million.

Obtaining a theoretical result for $a_\mu$ of comparable precision has proved a
more difficult task than in the case of $a_e$ \cite{Jegerlehner:2009ry}. This is
because, as stated above, the contributions from other sectors of the standard
model are more significant. However the calculation has been brought to a point
where the uncertainty is of the same order as the experimental uncertainty.
Interestingly however, there is a discrepancy between the two values which
exceeds the current uncertainty. This has attracted a huge amount of interest to
$a_\mu$ and lead to significant efforts to calculate contributions from
potential new-physics sectors.

The current uncertainty in $a_\mu$ is strongly dominated by hadronic contributions,
specifically the leading order hadronic, and hadronic light-by-light
contributions. The light-by-light contribution has attracted significant
theoretical interest, and has recently become the focus of considerable work
using lattice simulations \cite{Hayakawa:2005eq,Blum:2009zz}. 

This work involves the leading order hadronic contribution, which we denote as $a_\mu^{(2)had}$, the best estimate of
which is currently obtained by relating the hadronic vacuum polarisation of the
photon to the cross section for $e^+ e^-$ decay into hadrons, allowing a
dispersive integral over experimental data for the cross section \cite{Jegerlehner:2008zz}. 

Despite the apparent accuracy of the results obtained from this procedure,
there remain discrepancies between results from different data sets, and these discrepancies carry over to the total result for $a_\mu^{(2)had}$, and depending
on the choice of method can result in a $\sim 3\sigma$ or $\sim 1\sigma$ discrepancy between the theoretical and experimental result for $a_\mu$. As a result,
it is not clear if this method of obtaining the vacuum polarisation is under
good control \cite{Jegerlehner:2009ry, Jegerlehner:2008zz}. In addition, the discrepancy of $a_\mu$ is an important input for new-physics model builders when
constraining their models. As such an independent first-principles calculation of $a_\mu^{(2)had}$ could have a large impact on efforts to discern the nature
of physics at and above the electro-weak scale.

Attempts have been made to estimate this quantity using models of low energy QCD
\cite{deRafael:1993za}, however for a truly robust first-principles evaluation of this quantity we must turn to lattice field theory techniques.

This quantity was first tackled through lattice computation in quenched
simulations first with domain wall fermions \cite{Blum:2002ii} , followed
by a calculation with improved Wilson fermions \cite{Gockeler:2003cw}. The first
dynamical simulation followed \cite{Blum:2003se,Aubin:2006xv} using 2+1 flavour staggered
quarks, and several studies of this quantity are ongoing, using 2 flavours of improved Wilson fermions \cite{DellaMorte:2010sw} and twisted mass fermions
\cite{Feng:2011zk}. We present a calculation of $a_\mu^{(2)had}$ from a dynamical simulation of 2+1
flavour QCD with domain wall fermions. 

\section{Background}

The Land\'{e} $g$-factor of a fermion can be expressed in terms of the electromagnetic form factors $F_1$ and $F_2$ as
\begin{equation}
g=2\left[F_1(0)+F_2(0)\right].
\end{equation}
These form factors are defined in the effective electromagnetic scattering vertex whereby the expression for the tree-level graph
\\\newline
\begin{equation} 
\parbox{25mm}{
\begin{fmffile}{ammm1}
\begin{fmfgraph*}(60,60)
\fmftop{in}
\fmfbottom{up,down}
\fmf{photon,tension=2}{in,v1}
\fmf{fermion}{up,v1}
\fmf{fermion}{v1,down}
\fmflabel{$q,\,\mu$}{in}
\fmflabel{$p$}{up}
\fmflabel{$p^\prime$}{down}
\end{fmfgraph*}
\end{fmffile}
}=-ie\gamma_\mu
\end{equation}
\\
is replaced by its equivalent including all quantum corrections
\\\newline
\begin{equation} 
\parbox{25mm}{
\begin{fmffile}{ammm2}
\begin{fmfgraph*}(60,60)
\fmftop{in}
\fmfbottom{up,down}
\fmf{photon,tension=2}{in,v1}
\fmfblob{100}{v1}
\fmf{fermion}{up,v1}
\fmf{fermion}{v1,down}
\fmflabel{$q,\,\mu$}{in}
\fmflabel{$p$}{up}
\fmflabel{$p^\prime$}{down}

\end{fmfgraph*}
\end{fmffile}
}=-ie\Gamma_\mu(p^\prime,p)\equiv-ie\left[\gamma_\mu F_1(q^2)+\frac{i\sigma^{\mu\nu}q_\nu}{2m}F_2(q^2)\right].\label{formfac}
\end{equation}
\\
From the Born approximation it can be seen that $F_1(0)=1$ to all orders, and so
\begin{equation}
 a=\frac{g-2}{2}=F_2(0).
\end{equation}
We seek to compute the effect of hadronic vacuum polarisation contributions to $a_\mu$ which are obtained by calculating contributions to the graph in (\ref{formfac}) of the form
\\\newline
\begin{equation}
\parbox{25mm}{
\begin{fmffile}{ammm3}
\begin{fmfgraph*}(60,60)
\fmftop{in}
\fmfbottom{up,down}
\fmf{photon,tension=1}{in,v1}
\fmf{fermion,tension=.5}{v2,v1}
\fmf{fermion,tension=.5}{v1,v3}
\fmf{photon,tension=0.15}{v2,v4,v3}
\fmfv{label=had,label.dist=0,decor.size=180,decor.shape=circle,decor.filled=empty}{v4}
\fmf{fermion}{up,v2}
\fmf{fermion}{v3,down}
\fmflabel{$q,\,\mu$}{in}
\fmflabel{$p$}{up}
\fmflabel{$p^\prime$}{down}
\end{fmfgraph*}
\end{fmffile}
}.\label{hadvacpol}
\end{equation}
\\
As described in \cite{Blum:2002ii} the contribution to $a_\mu$ from the one-loop diagram equivalent to the graph (\ref{hadvacpol}) with the hadronic blob removed can be expressed as
\begin{equation}
\parbox{25mm}{
\begin{fmffile}{ammm4}
\begin{fmfgraph*}(60,60)
\fmftop{in}
\fmfbottom{up,down}
\fmf{photon,tension=1}{in,v1}
\fmf{fermion,tension=0.5}{v2,v1}
\fmf{fermion,tension=0.5}{v1,v3}
\fmf{photon,tension=0.03}{v2,v3}
\fmf{fermion,tension=0.5}{up,v2}
\fmf{fermion,tension=0.5}{v3,down}
\end{fmfgraph*}
\end{fmffile}
}\longrightarrow a_\mu^{(1)}=\frac{\alpha}{\pi}\int_0^\infty dQ^2\,f(Q^2)
\end{equation}
 where the kernel function $f(Q^2)$ is divergent as $Q^2\rightarrow0$ and can be expressed
\\
\begin{eqnarray}
 f(Q^2)=\frac{m_\mu^2Q^2Z(Q^2)^3(1-Q^2Z(Q^2))}{1+m_\mu^2Q^2Z(Q^2)^2}
& &Z(Q^2)=-\frac{Q^2-\sqrt{Q^4+4m_\mu^2Q^2}}{2m_\mu^2Q^2}.
\end{eqnarray}
\\
From this, the expression for the hadronic vacuum polarisation contribution can be obtained with the insertions:
\\
\begin{equation}
\parbox{25mm}{
\begin{fmffile}{ammm5}
\begin{fmfgraph*}(60,60)
\fmftop{in}
\fmfbottom{up,down}
\fmf{photon,tension=1}{in,v1}
\fmf{fermion,tension=.5}{v2,v1}
\fmf{fermion,tension=.5}{v1,v3}
\fmf{photon,tension=0.15}{v2,v4,v3}
\fmfv{label=had,label.dist=0,decor.size=180,decor.shape=circle,decor.filled=empty}{v4}
\fmf{fermion}{up,v2}
\fmf{fermion}{v3,down}
\end{fmfgraph*}
\end{fmffile}
}\longrightarrow a_\mu^{(2)had}=\left(\frac{\alpha}{\pi}\right)^2\int_0^\infty dQ^2\,f(Q^2)\times\hat{\Pi}(Q^2)
\label{hadvacint}
\end{equation}
where $\hat{\Pi}(Q^2)$ is the infra-red subtracted transverse part of the hadronic vacuum polarisation
\begin{align}
\hat{\Pi}(Q^2)=\Pi(Q^2)-\Pi(0)&&\Pi_{\mu\nu}(q)=(q^2g_{\mu\nu}-q_\mu q_\nu)\Pi(q^2)\label{transverse}
\end{align}
\begin{equation}
\parbox{30mm}{
\begin{fmffile}{ammm6}
\begin{fmfgraph*}(60,60)
\fmfleft{in}
\fmfright{out}
\fmf{photon,tension=1}{in,v1}
\fmf{photon,tension=1}{v1,out}
\fmfv{label=had,label.dist=0,decor.size=180,decor.shape=circle,decor.filled=empty}{v1}
\fmflabel{$q,\,\mu$}{in}
\fmflabel{$q,\,\nu$}{out}
\end{fmfgraph*}
\end{fmffile}
}\equiv i\Pi_{\mu\nu}(q)
\end{equation}
at Euclidean momentum $Q^2=-q^2$. The hadronic vacuum polarisation function $\Pi_{\mu\nu}(q)$ can be computed as the Fourier-transformed two-point correlator
\begin{equation}
\Pi_{\mu\nu}(q)=\int d^4x \,e^{iq\cdot (x-y)}\langle J_\mu(x)
J_\nu(y)\rangle\label{vacpolcorr}
\end{equation}
involving the electromagnetic current
\begin{equation}
 J_\mu(x)=\sum_iQ_i\bar{\psi}^i\gamma_\mu\psi^i
\end{equation}
where $\psi^i$ is the quark field of flavour $i$ and $Q^i$ is its charge. The path-integral used in the expectation value in (\ref{vacpolcorr}) will involve only hadronic fields, i.e. quarks and gluons. 

\subsection{Simulation}
Our computation is performed using configurations generated by the RBC \& UKQCD
collaborations as part of their program of investigation using 2+1 flavours of domain-wall fermions. We investigate three lattice volumes, each with several ensembles at different values of the light quark
mass $m_u$. The parameters of these ensembles are given in Table \ref{tab:param}. The ensembles at $\beta=1.75$ have been generated using a dislocation suppressing
determinant ratio (DSDR) in conjunction with the Iwasaki gauge action, with a fifth dimension whose extent is $L_5$=32 \cite{Ohta:2011nv,Kelly:2011up}. The lighter of
these ensembles is very near to the physical point with a pion mass of $m_\pi\simeq 180$ MeV. The other ensembles used only the Iwasaki action and $L_5=16$ \cite{Allton:2008pn,Aoki:2010dy}.

\begin{table}[!htp]
\centering
\begin{tabular}{c|c|c|c|c|c}
$V$ & $\beta$ & $a^{-1} $ GeV & $\hat{q}^2_{min}$ GeV$^2$ & $am_h$ & $am_u$ \\\hline\hline
$24^3\times64$ & 2.13 & 1.73(2) & 0.028 & 0.04 & 0.02  \\
$24^3\times64$ & 2.13 & 1.73(2) & 0.028 & 0.04 & 0.01 \\
$24^3\times64$ & 2.13 & 1.73(2) & 0.028 & 0.04 & 0.005 \\
$32^3\times64$ & 2.25 & 2.28(3) & 0.05 & 0.03 & 0.008 \\
$32^3\times64$ & 2.25 & 2.28(3) & 0.05 & 0.03 & 0.006\\
$32^3\times64$ & 2.25 & 2.28(3) & 0.05 & 0.03 & 0.004 \\
$32^3\times64$ & 1.75 & 1.375(9) & 0.018 & 0.045 & 0.0042\\
$32^3\times64$ & 1.75 & 1.375(9) & 0.018 & 0.045 & 0.001 
\end{tabular}
\caption{\small Parameters of the lattice ensembles used in our study.}
\label{tab:param}
\end{table}

\begin{table}[!htp]
\centering
\begin{tabular}{c|c|c|c|c|c}
 $\beta$ & $am_u$ & $Z_V$ &$am_\mathrm{V}$&$am_\mathrm{PS}$ & $af_\mathrm{V}$ \\\hline\hline
 2.13 & 0.02 & 0.696(2)&0.579(6)&0.3227(7) & \\
 2.13 & 0.01 & 0.700(2)& 0.529(5)&0.2422(5)& \\
 2.13 & 0.005 & 0.699(2) & 0.505(6)&0.1904(6)& \\
 2.25 & 0.008 & 0.7380(5) & 0.388(6)&0.1727(4)& 0.078(6)\\
 2.25 & 0.006 & 0.7385(6)& 0.366(5)&0.1512(3)& 0.076(5)\\
 2.25 & 0.004 & 0.7387(7)  & 0.356(6)&0.1269(4) & 0.070(11)\\
 1.75 &  0.0042 & 0.664(5)& 0.570(25) & 0.1809(3)& 0.102(6)\\
 1.75 &  0.001 & 0.669(8) & 0.558(44) & 0.1249(3)& 0.105(15)
\end{tabular}
\caption{\small Relevant observables measured on our lattices. Results on the $\beta=1.75$ lattices are preliminary and will be outlined in a forthcoming publication \cite{Kelly:2011up}, results for $f_\mathrm{V}$ on the $64\times 24^3$ lattices are currently unavailable.}
\label{tab:obsv}
\end{table}

\subsection{Vacuum polarisation}
\label{vacpol}
We compute the lattice vacuum polarisation as
\begin{equation}
 \widetilde{\Pi}_{\mu\nu}(x)=Z_V\sum_i 
Q_i^2  a^6\langle 
\mathcal{V}^i_\mu(x)V^i_\nu(0)\rangle,
\end{equation}
where we have omitted the flavour-nondiagonal terms as they contain only ``disconnected'' contributions which are expected to be sub-dominant, as will be discussed further below.

At the sink we use the DWF conserved vector current \cite{Furman:1994ky}
\begin{equation}
 \mathcal{V}^i_\mu(x)=\sum_{s=1}^{L_5}\frac{1}{2}\left[\bar{\psi}^i(x+\hat{\mu},s)(1+\gamma_\mu)U_\mu^\dagger(x)\psi^i(x,s)-\bar{\psi}^i(x,
s)(1-\gamma_\mu)U_\mu(x)\psi^i(x+\hat{\mu},s)\right]
\end{equation}
while at the source we have the local vector current
 $V^i_\nu(x)=\bar{q}^i(x)\gamma_\nu q^i(x)$
where $q^i(x)=P_+\psi^i(x,L_5-1)+P_-\psi^i(x,0)$, and $P_\pm=\frac{1}{2}(1\pm\gamma_5)$. Because of the use of the local vector current, a factor of the vector current renormalisation constant, $Z_V$, is included in our definition of the vacuum polarisation. The values of $Z_V$ used on each ensemble are given in Table \ref{tab:obsv}, as measured in \cite{Aoki:2010dy}.


These correlators were generated for, and used in, the measurement of the QCD contribution to the electro-weak S-parameter \cite{Boyle:2009xi}. However, they will prove perfectly sufficient for our purposes, as long as we are mindful of Ward Identity violations, which will be discussed in Sec. \ref{sec:ward}. 

Of the two Wick-contractions arising from this correlator, we compute only the connected one. We leave the evaluation of the disconnected contribution for
future work, but note that it is expected to be suppressed relative to the connected contribution \cite{DellaMorte:2010aq}. This argument is also the motivation for neglecting the flavour-nondiagonal terms, and we will make an estimate of the systematic uncertainty that results in our conclusions.

We Fourier transform into momentum space: 
\begin{equation}
 \widetilde{\Pi}_{\mu\nu}(\hat{q})\equiv Z_V\sum_i Q_i^2\sum_x e^{iqx}a^6\langle 
\mathcal{V}^i_\mu(x)V^i_\nu(0)\rangle\label{correlator}
\end{equation}
using the discrete momenta 
 $q_\mu=\frac{2\pi n_\mu}{L_\mu}$
where $n_\mu$ is a 4-tuple of integers, and $L_\mu$ is the length of the lattice in the $\mu$ direction. From here, we will use the lattice momentum
\begin{equation}
 \hat{q}_\mu=\frac{2}{a}\sin\left(\frac{\pi n_\mu}{L_\mu}\right).
\end{equation}
We associate the quantity $\hat{q}^2=\sum_\mu\hat{q}_\mu^2$ with the continuum momentum $Q^2$.



\subsection{Ward identities}
\label{sec:ward}
In order to ensure that this reproduces a vacuum polarisation of the form
(\ref{transverse}) we must verify that this lattice correlator satisfies the Ward
identity
$q_\mu\Pi_{\mu\nu}=0$ which in general is not the case, as although both
operators $\mathcal{V}^i$ and $V^i$ have the correct continuum limit 
\begin{equation}
 \mathcal{V}_\mu^i,V_\mu^i\stackrel{a\rightarrow0}{\longrightarrow}J^i=\bar{\psi}^i\gamma_\mu\psi^i
\end{equation}
the additional irrelevant operators introduced into the lattice action modify the
Ward identity for $\widetilde{\Pi}_{\mu\nu}$.
In coordinate space, the Schwinger Dyson equation for
$\widetilde{\Pi}_{\mu\nu}$ reads
\begin{equation}
 \langle(\Delta_\mu
\mathcal{V}^i_\mu(x))V_\nu^i(0)\rangle+\left\langle\left(\frac{
V_\nu^i(0)\overleftarrow{\partial}}{\partial\psi^i(x)}
\psi^i(x)\right)-\left(\bar{\psi}^i(x)\frac
{\vec{\partial}V_\nu^i(0)}{\partial\bar{\psi}^i(x)}\right)\right\rangle=0
\label{sdeq}
\end{equation}
where $\Delta_\mu$ is the backward lattice derivative. Because the local current
used is not point-split, the second term in (\ref{sdeq}) vanishes and we have
as a result that 
$e^{\frac{iaq_\mu}{2}}\hat{q}_\mu\widetilde{\Pi}_{\mu\nu}=0$.

\begin{figure}[!htp]
\subfloat[$a\hat{q}_\mu\widetilde{\Pi}_{\mu\nu}$ ]{
\includegraphics[scale=0.3]{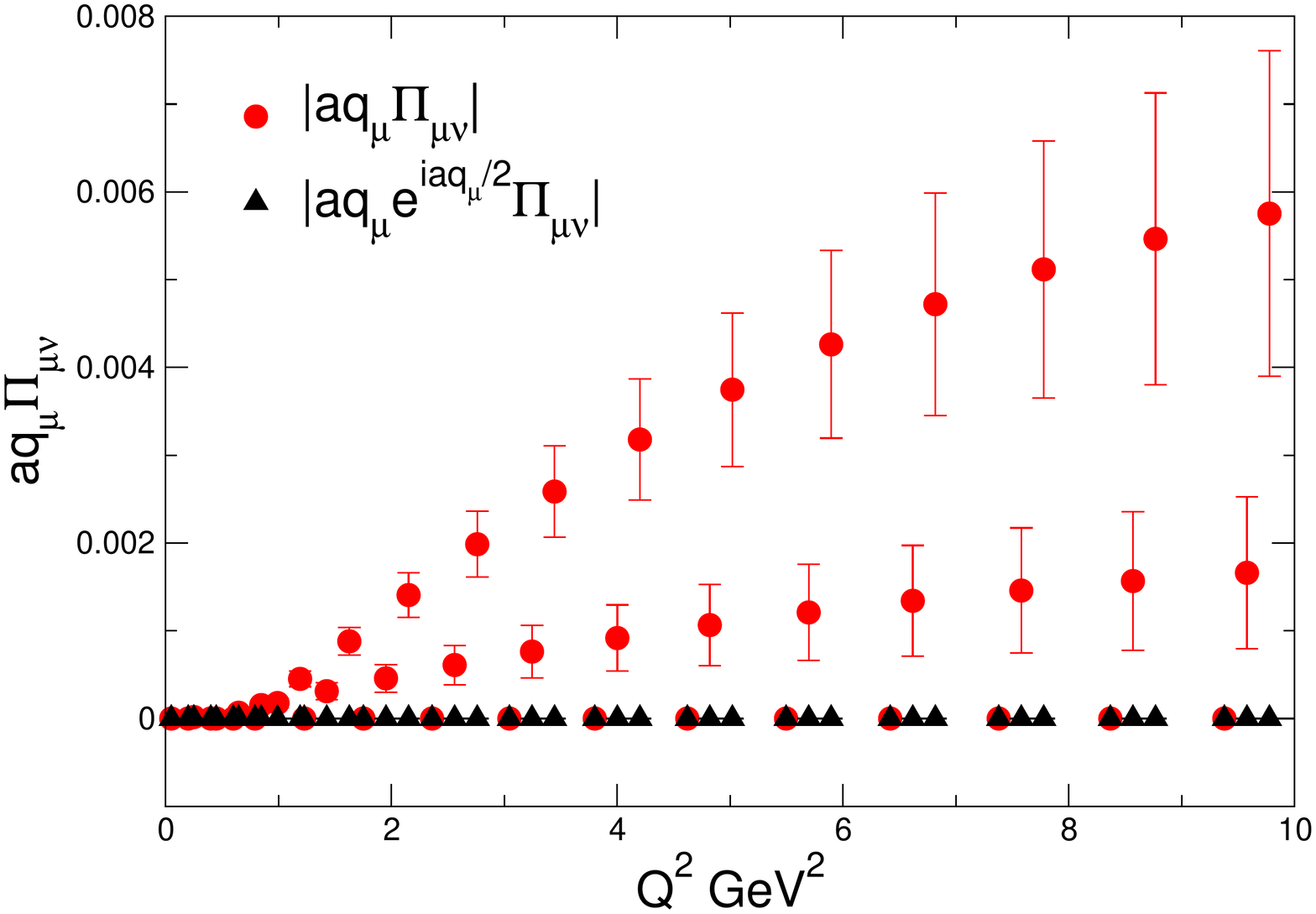}
\label{disc2mpcac}
}
\subfloat[$\widetilde{\Pi}_{\mu\nu}a\hat{q}_\nu$]{
\includegraphics[scale=0.3]{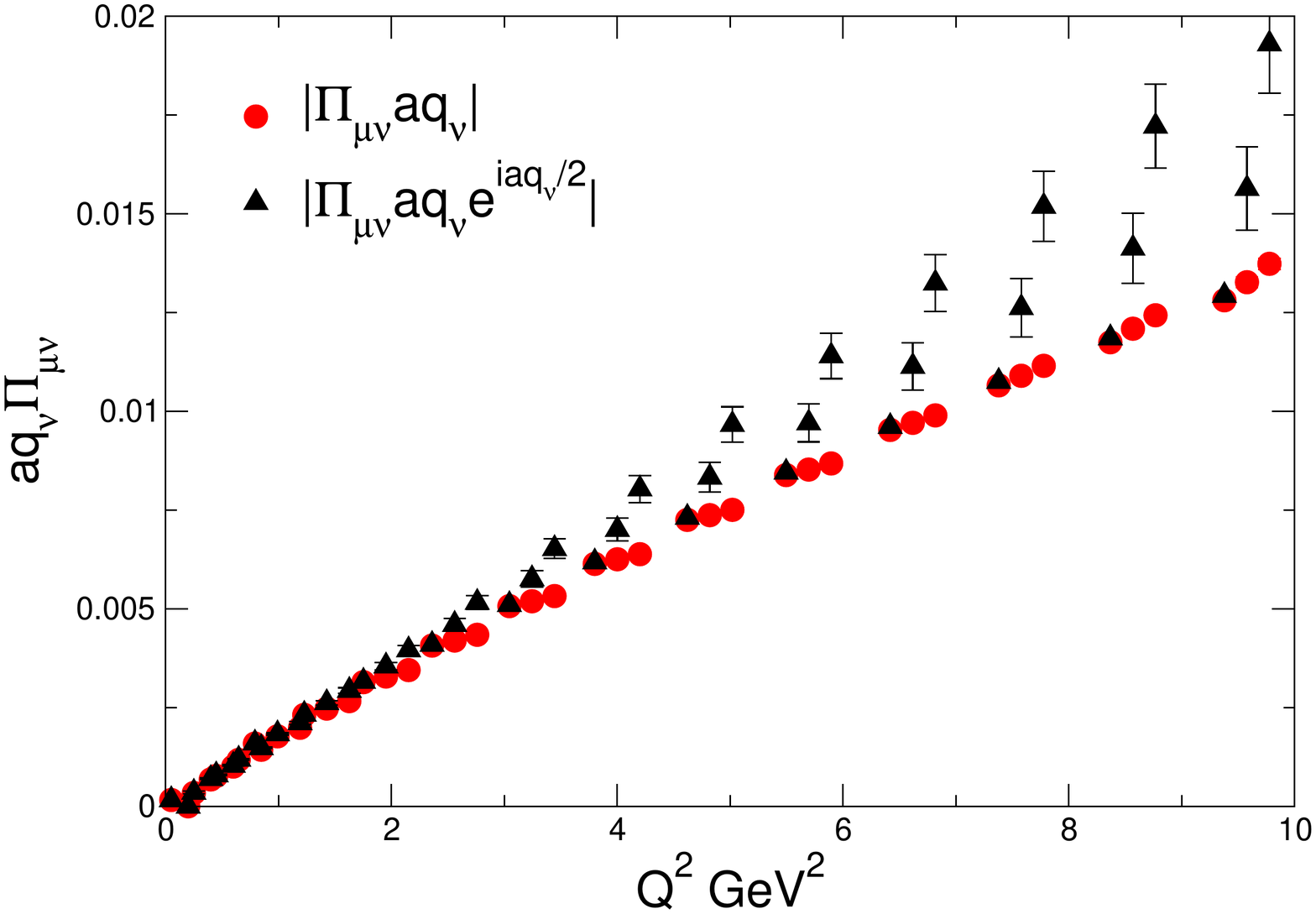}
\label{disc2mps}
}
\caption{\small Illustration of Ward identity violations in $\Pi_{\mu\nu}$ on $32^3\times64$ lattice at $\beta=2.25$ and $am_u=0.004$.}\label{wardplots}
\end{figure}

This is illustrated in Fig. \ref{wardplots} where we see that it is necessary to
include the factor $e^{i\frac{aq_\mu}{2}}$ in the Ward identity for the first
index of $\widetilde{\Pi}_{\mu\nu}$, while there is no fulfilled Ward identity
for the second index.

\subsection{Decomposing the vacuum polarisation}
\label{decomp}

We must extract from $\widetilde{\Pi}_{\mu\nu}(\hat{q})$ the scalar vacuum
polarisation $\widetilde{\Pi}(\hat{q}^2)$ which, corresponding to the
continuum
(\ref{transverse}), are related by

\begin{equation}
 \widetilde{\Pi}_{\mu\nu}(\hat{q})=(\hat{q}^2\delta_{\mu\nu}-\hat{q}_\mu\hat{q
}_\nu)\widetilde{\Pi}(\hat{q}^2)
\end{equation}

In practice, in order to avoid any longitudinal contribution which might arise
due to the non-conservation of Ward identities, for each momentum orientation we
choose directions $\mu$ such that $\hat{q}_\mu=0$ and compute  

\begin{equation}
 \widetilde{\Pi}(\hat{q}^2)=\frac{\widetilde{\Pi}_{\mu\mu}(\hat{q})}{\hat{q}
^2}
\end{equation}
where in the above there is no sum over $\mu$. 

\begin{figure}[!htp]
\includegraphics[scale=0.5]{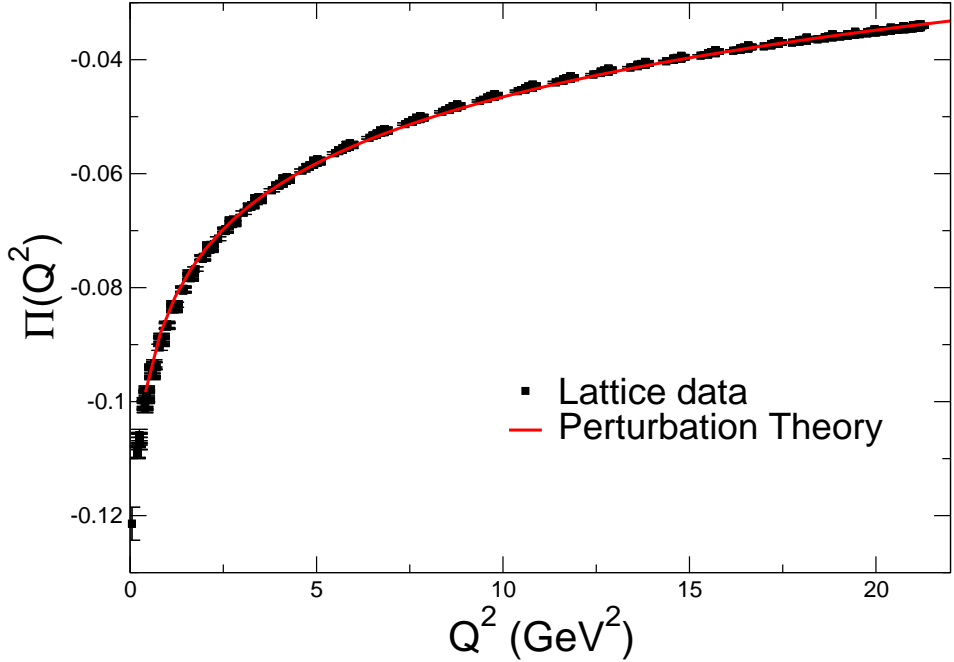}
\caption{\small Vacuum polarisation function $\Pi(Q^2)$ as measured on $64\times32^3$ lattice at $\beta=2.25$ and $am_u=0.004$.}\label{platplots}
\end{figure}

In Fig. \ref{platplots} we show an example of the resulting vacuum polarisation function, and compare this to the large $Q^2$ expansion of the three-loop
continuum perturbation theory result from \cite{Chetyrkin:1996cf}, using two massless flavours of quarks and one massive flavour which we associate with the
strange quark. This result is quoted in the $\overline{\mathrm{MS}}$ scheme and as such we require the strange quark mass in our simulations expressed in
$\overline{\mathrm{MS}}$. For this we use the non-perturbative renormalization factor $Z^{\overline{\mathrm{MS}}}_{mh}=0.1533(6)(33)$ determined in
\cite{Aoki:2010dy}. The factor is quoted in the limit of vanishing light quark mass, but it is also illustrated that the mass dependence is extremely slight,
and so we see this as satisfactory.

\section{Deducing $a^{(2)had}_\mu$}

In order to infer the value of $a^{(2)had}_\mu$ from our data we must carry out
the integral (\ref{hadvacint}) which we split into high and low momentum regions at some momentum cut $Q_C^2$ 

\begin{equation}
 a_\mu^{(2)had}=4\alpha^2\left[\int_0^{Q^2_C}\,dQ^2f(Q^2)\times\hat{\Pi}
(Q^2)+\int_{Q^2_C}^\infty\,dQ^2f(Q^2)\times\hat{\Pi}(Q^2)\right]\label{eq:integral}.
\end{equation}

A continuous description of $\Pi(Q^2)$ at low momenta is obtained by performing a fit to our lattice data, which allows us to perform the low $Q^2$ integral.
The value of $\Pi(0)$ from this fit combined with a high-momentum description of $\Pi(Q^2)$ from perturbation theory allows us to perform the high momentum
integral. As we shall see, the integral is strongly dominated by the low momentum contribution.

\subsection{Fitting the low $Q^2$ region}

We have attempted to fit a continuous form to our lattice data for the vacuum polarisation using a number of different fit forms. The effect that the choice of fit function can have on the result for $a_\mu^{(2)had}$ has been highlighted in previous studies \cite{Aubin:2006xv}, and this behoves us to ensure that the systematics with regard to this choice are under control.

The suitability of a given fit-form should be judged on two main criteria: 
\begin{itemize}
 \item Firstly, the chosen expression must describe the data closely, and must do so regardless of the range of data included in the fit. As such we require the reduced $\chi^2$ of the fit to be consistently low as a function of $Q_C^2$ which defines the range of data in the fit. 
\item Secondly, in order to deduce that the fit-form results in an integral over momentum which is relatively stable, we desire that the result for $a_\mu^{(2)had}$ is again relatively stable as a function of $Q_C^2$. 
\end{itemize}

Ref. \cite{Aubin:2006xv} also illustrated the use of a fit form originating in the expression for the vacuum polarisation calculated in chiral perturbation theory. The dominant component of this expression is due to the vector meson contribution, which at tree-level is
\begin{equation}
 \Pi_V^{tree}(Q^2)=\frac{2}{3}\frac{f_V^2}{Q^2+m_V^2}\label{eq:vecpole}
\end{equation}
where the vector decay constant $f_V$ is defined
\begin{equation}
 \langle\Omega|J_\mu|V,p,\epsilon\rangle=m_Vf_V\epsilon_\mu(p).
\end{equation}

Motivated by this expression the fit-form we use is closely related, differing only in the inclusion of the contribution of an additional vector resonance,
\begin{equation}
 \Pi(Q^2)=A-\frac{F_1^2}{Q^2+m_1^2}-\frac{F_2^2}{Q^2+m_2^2}\label{eq:twovec}.
\end{equation}
The one-loop contribution from the pseudoscalar sector, shown in \cite{Aubin:2006xv} to have small momentum dependence, will not strongly affect our results
and so, in our effort to make a continuous description of the lattice data, it will be omitted from our fit ansatz. 

\begin{figure}[!htp]
\centering
\includegraphics[scale=0.5]{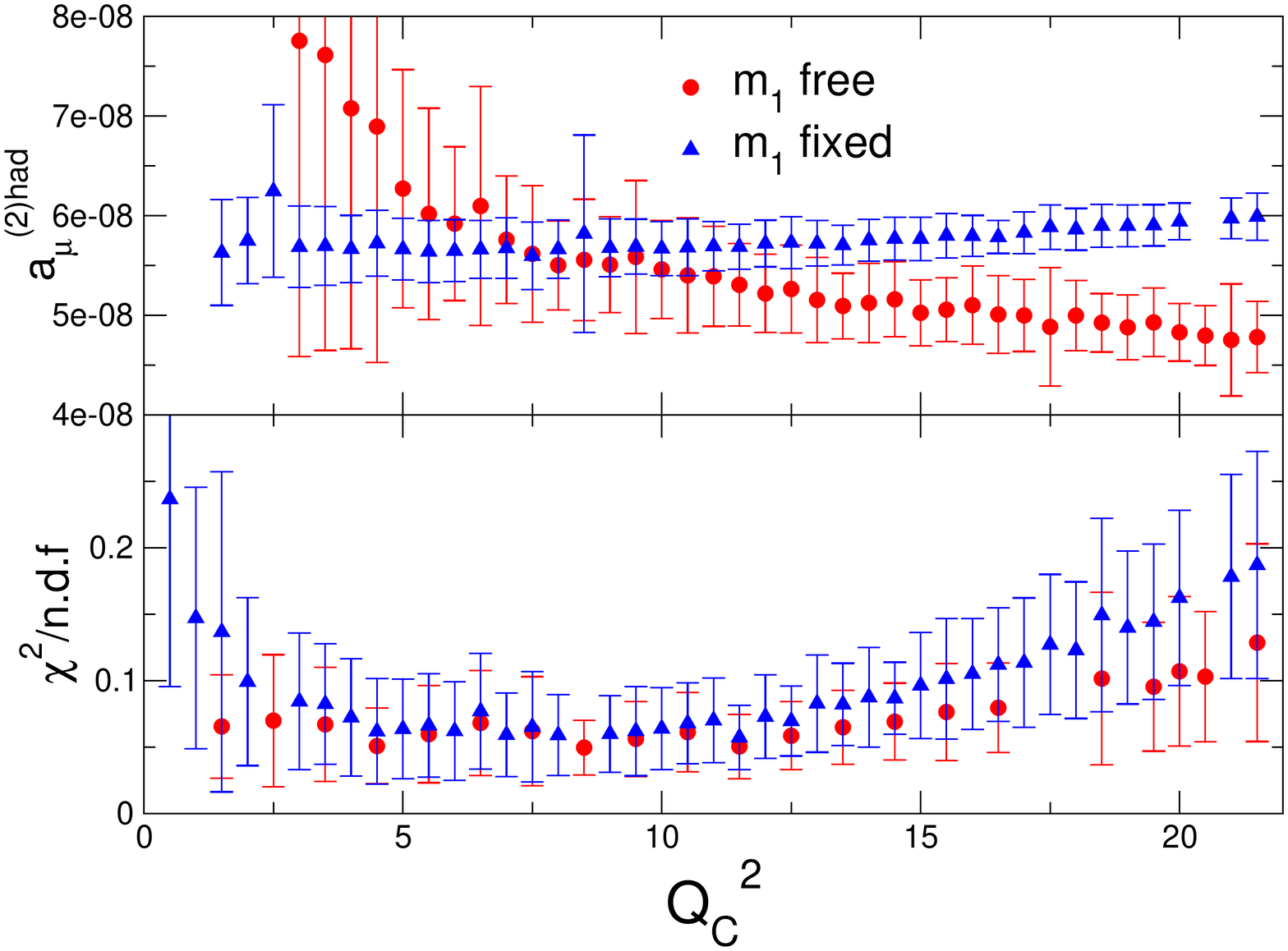}
\caption{\small Properties of fits to the lattice vacuum polarisation using the ansatz (\ref{eq:twovec}) on the $\beta=2.25$ lattice at $am_u=0.004$.}
\label{fig:twovec}
\end{figure}

We fit the lattice vacuum-polarisation data in two ways:
\begin{itemize}
 \item Firstly using $A$, $F_{1,2}$ and $m_{1,2}$ as free parameters.
\item Also, fixing the parameter $m_1$ to the mass of the vector meson $m_\mathrm{V}$ as measured in \cite{Aoki:2010dy}. This we do by constraining $m_1$ to lie in the one-sigma band defined by the estimate of $m_\mathrm{V}$ and its variance. This method was found to maintain the stability of the fit routine, while incorporating the extra information provided by $m_\mathrm{V}$. In this fit $A$, $F_{1,2}$ and $m_2$ remain as true free parameters.
\end{itemize}
The behaviour of such fits are shown in Fig. \ref{fig:twovec}. 
Clearly such a form is a very good representation of the data, over practically the whole range of $Q_C^2$. In addition the results for $a_\mu^{(2)had}$
using such fits are very stable as the fit range is varied, allowing far greater confidence in the reliability of the result. In particular we conclude
that using a fit form (\ref{eq:twovec}) with the mass of the first pole fixed to the ground-state vector meson mass to be the optimal method of describing
the lattice data for the hadronic vacuum polarisation. 

\begin{figure}[!htp]
\centering
\includegraphics[scale=0.5]{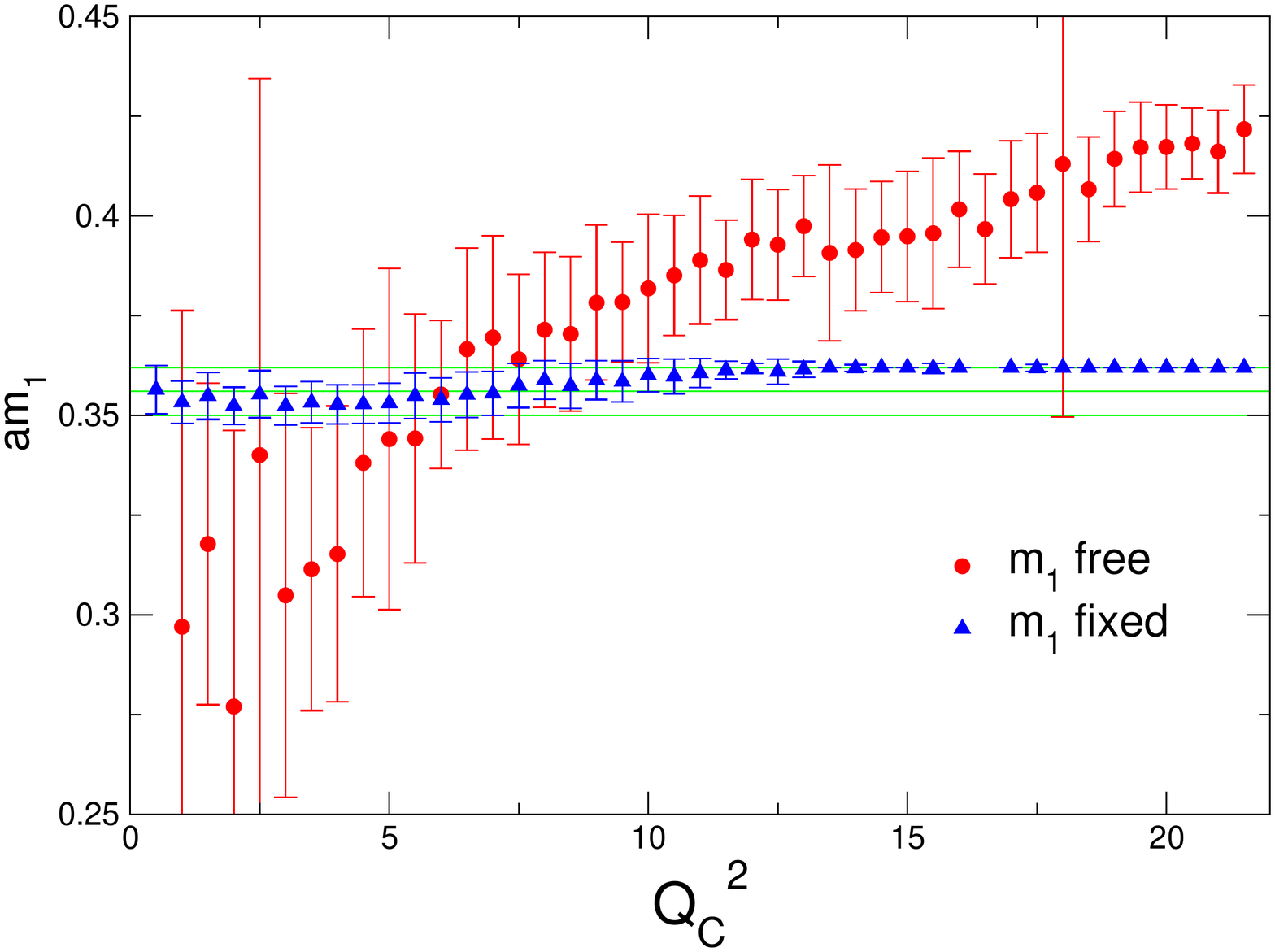}
\caption{\small Value of the fit parameter $am_1$ in fits using the ansatz (\ref{eq:twovec}) on the $\beta=2.25$ lattice at $am_u=0.004$. The vector mass $am_\mathrm{V}$ as
determined on this lattice is shown in green. Note in the fit where $m_1$ was fixed, it was only constrained to lie within the green band. It is clear that for
a high $Q_C^2$, $m_1$ will emerge at the upper limit of the band, indicating some tension between the fit-form and the data, but as can be seen in Fig.
\ref{fig:twovec}, this has very little impact on the goodness of the fit.}
\label{fig:mv}
\end{figure}

In Fig. \ref{fig:mv} we see the value of the fit parameter $m_1$ from (\ref{eq:twovec}) as determined from fits to the lattice vacuum polarisation. The value of $m_\mathrm{V}$ obtained in \cite{Aoki:2010dy} is shown in green, and this defines the band in which $m_1$ was constrained to reside in the fixed version of this fit. 
We have not attempted to model $O(4)$ breaking effects present in our data. Though such effects do appear to be present to a moderate degree on certain
ensembles, they do not prevent the extraction of a reasonable signal from our data at this point. These effects could also be alleviated by the use of twisted boundary conditions \cite{Arthur:2010ht}.

\subsection{Evaluation of (\ref{eq:integral})}

\begin{figure}[!htp]
 \centering
\subfloat[
$\beta=2.25$  \hskip0.5cm$am_u=0.004$
\newline $Q_C^2=11$ GeV$^2$
]{
\includegraphics[scale=0.25]{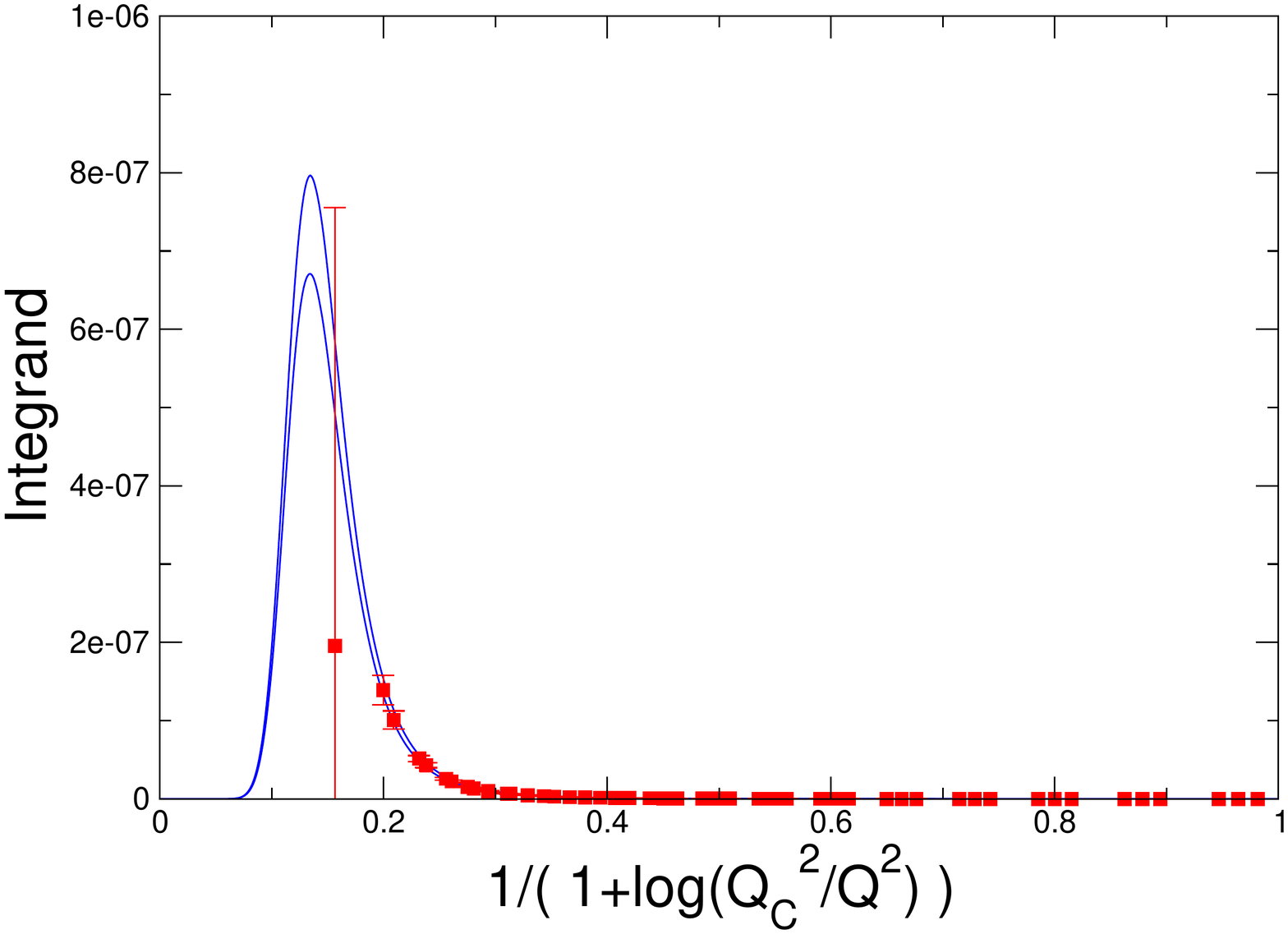}
\label{firstint}
}
\subfloat[$\beta=1.75$\hskip0.5cm$am_u=0.0042$ \newline$Q_C^2=4$ GeV$^2$]{
\includegraphics[scale=0.25]{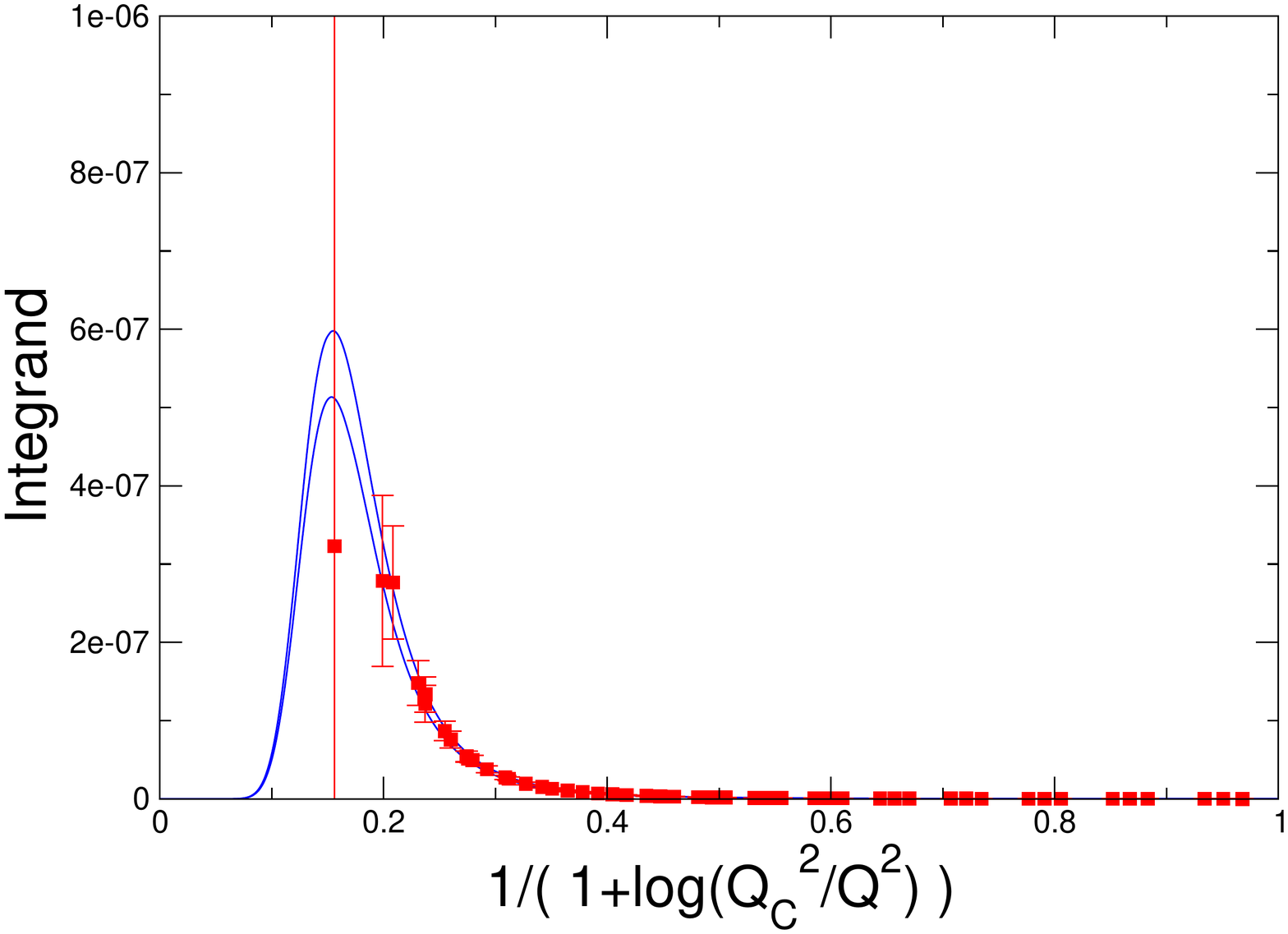}
\label{dsdrint}
}
\caption{\small Examples of the integrand in the rescaled integral (\ref{moment2}). }\label{fig:ints}
\end{figure}

Illustrations of the integrand can be seen in Fig. \ref{fig:ints}. 
Because the integrand is dominated by contributions in the low momentum region, we change our
integration measure to better sample the region of interest. To do this, we make the change of variables

\begin{equation}
t=\frac{1}{1+\log{\frac{Q_C^2}{Q^2}}}
\end{equation}
and so the integral over the low-momentum region becomes

\begin{equation}
\int_0^{Q_C^2}\,dQ^2f(Q^2)\times\hat{\Pi}(Q^2)\longrightarrow\int_0^1\,dt\,f(Q^2)\times\hat{\Pi}(Q^2)\times\frac{Q^2}{t^2}
\label{moment2}.
\end{equation}

Overlaid on the depiction of the integrand in Fig. \ref{fig:ints} is the appropriately subtracted and rescaled vacuum polarisation data. We see from this
that, while a large portion of the constraint on the fit is consistently derived from data at higher momentum, the fit is always consistent with the data at
low
momentum, the region where the integral receives the dominant contribution.

In particular in Fig.\,\ref{dsdrint} we see that on the larger lattices at $\beta=1.75$ using the Iwasaki+DSDR action, the data point at the lowest momentum
sits exactly where the integrand reaches a maximum, and there are numerous data points in the dominant region, constraining the fit. Clearly using lattices of
such size will help in obtaining a precise result for this quantity, and this must be combined with the use of twisted boundary conditions \cite{DellaMorte:2010sw} in order to access data at lower values of the lattice momentum.

\section{Results}

We extract our final results from the fit using (\ref{eq:twovec}) with the first mass fixed to that of the vector meson as measured on each ensemble. Observing the behaviour of the reduced
$\chi^2$ as the fit range is varied, we choose a suitable value for $Q_C^2$ for each
ensemble which provides the most reliable result. We attempt to choose a cut which provides a low reduced $\chi^2$ preferably where the parameter $m_1$ agrees without tension with $m_\mathrm{V}$. This produces the results shown in Table \ref{tab:results}, where we also quote the reduced $\chi^2$ of the
fit, and the resulting values of the remaining associated free parameters. 

\begin{figure}[!htp]
 \centering
\includegraphics[scale=0.5]{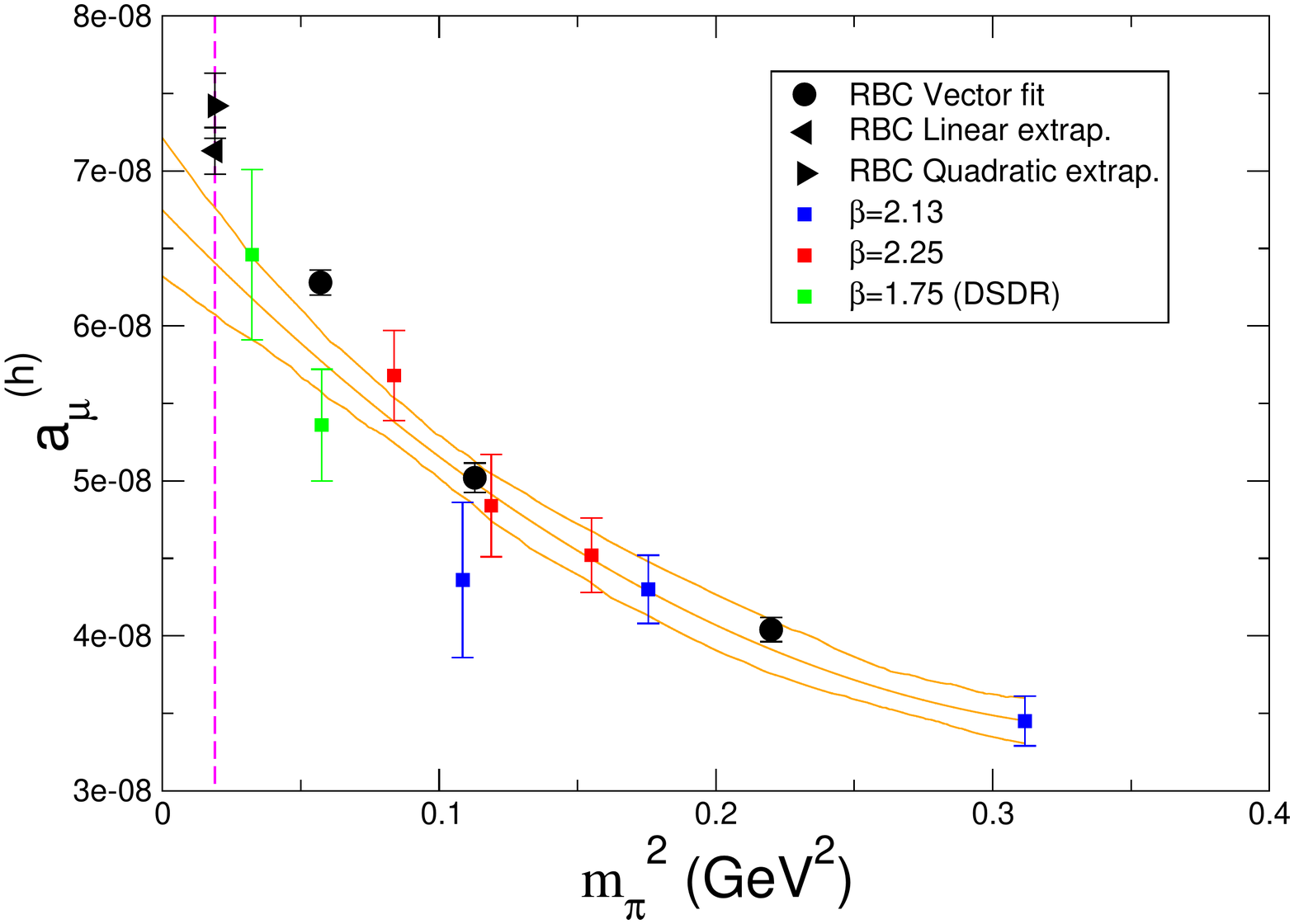}
\caption{\small Integrated result for $a_\mu^{(2)had}$ as a function of the pseudoscalar
mass squared.}\label{resplot}
\end{figure}

These results are also shown as a function of $m_\pi^2$ in Fig. \ref{resplot}, where we compare them to previous 2+1 flavour results from \cite{Aubin:2006xv}.
Also shown is an extrapolation to the physical point, using a quadratic chiral ansatz. This produces a final result for the leading order hadronic vacuum
polarisation contribution the anomalous magnetic moment of the muon 
\begin{equation}
 a_\mu^{(2)had}=641(33)\times 10^{-10}\label{eq:standres}.
\end{equation}

\begin{figure}[!htp]
 \centering
\subfloat[Results using modified prescription $H=m_V$.]{
\includegraphics[scale=0.27]{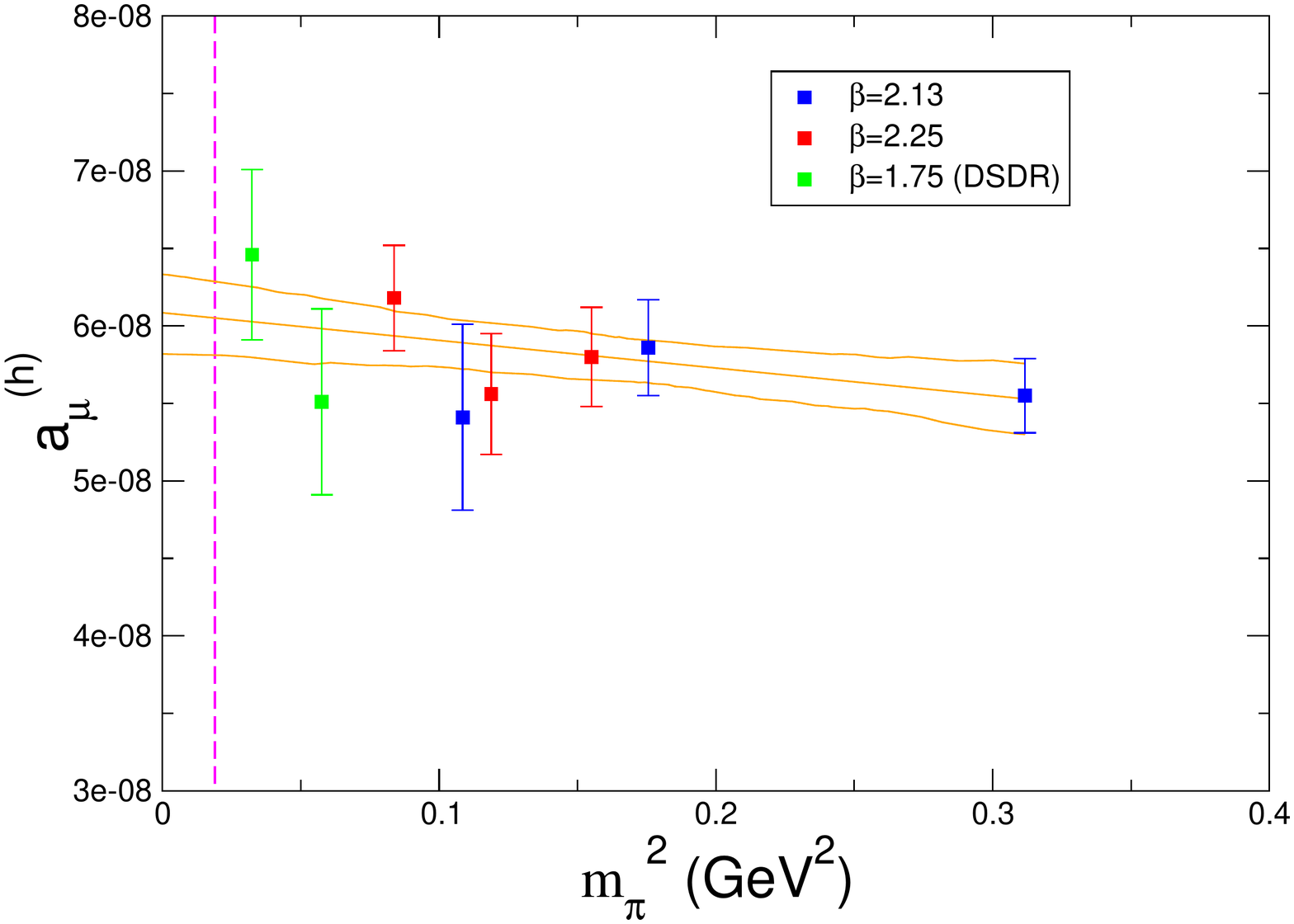}
\label{fig:jansena}
}
\subfloat[Comparison of results from both methods.]{
\includegraphics[scale=0.27]{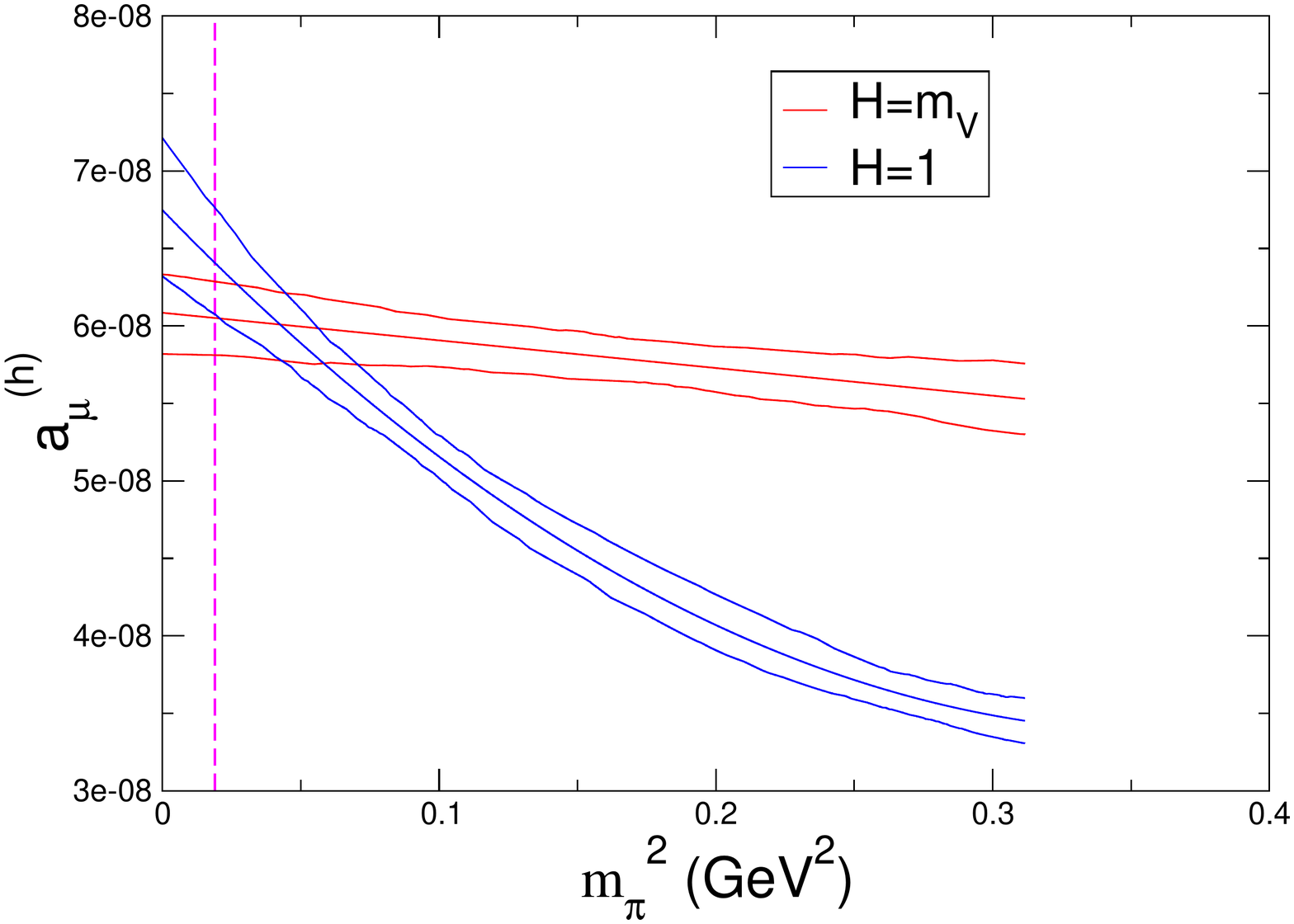}
\label{fig:jansenb}
}
\caption{\small Analysis of results for modified prescription using $H=m_V$.}
\end{figure}

We have also investigated the effect of modifying the kernel function in the integrand (\ref{hadvacint}) in the manner outlined in \cite{Feng:2011zk}, where in
an effort to moderate the variation of the outcome of the integral as a function of the quark mass, the momentum argument of the kernel function is rescaled by
a factor of the ratio of the value of a relevant observable $H$ (the mass of the vector meson appears to be an optimal choice) measured at the simulated quark mass to
its physical value. This effectively defines the calculation of a new quantity which approaches the desired $a_\mu^{(2)had}$ in the physical limit. We show the results of such a calculation in Fig. \ref{fig:jansena}, along with an accompanying chiral extrapolation. The chiral variation in this redefined quantity is such that it allows for a linear extrapolation in quark mass. For the lightest point in our simulation we include the unmodified result outlined in Table \ref{tab:results} since for this ensemble the measured vector mass $m_V$ is consistent with the physical value. This method does indeed moderate the chiral behaviour of the result, however it has little effect on our data at light quark masses, primarily because the lattice vector meson masses are very near that of the physical $\rho$ meson, and, as of now, are not determined to any great precision on these lattices. As such
this technique does not improve our chiral fit at this time, producing a compatible result with a similar uncertainty:
\begin{equation}
 a_\mu^{(2)had}=605(24)\times 10^{-10}\label{eq:janres}
\end{equation}
In Fig.\,\ref{fig:jansenb} we compare both chiral extrapolations, with $H=1$ denoting the standard method, and $H=m_V$ indicating the modified prescription of \cite{Feng:2011zk} using the vector mass $m_V$.

\begin{figure}[!htp]
 \centering
\subfloat[Hadronic vacuum polarisation contribution $a_\mu^{(2)had}$ to $a_\mu$.]{
\includegraphics[scale=0.27]{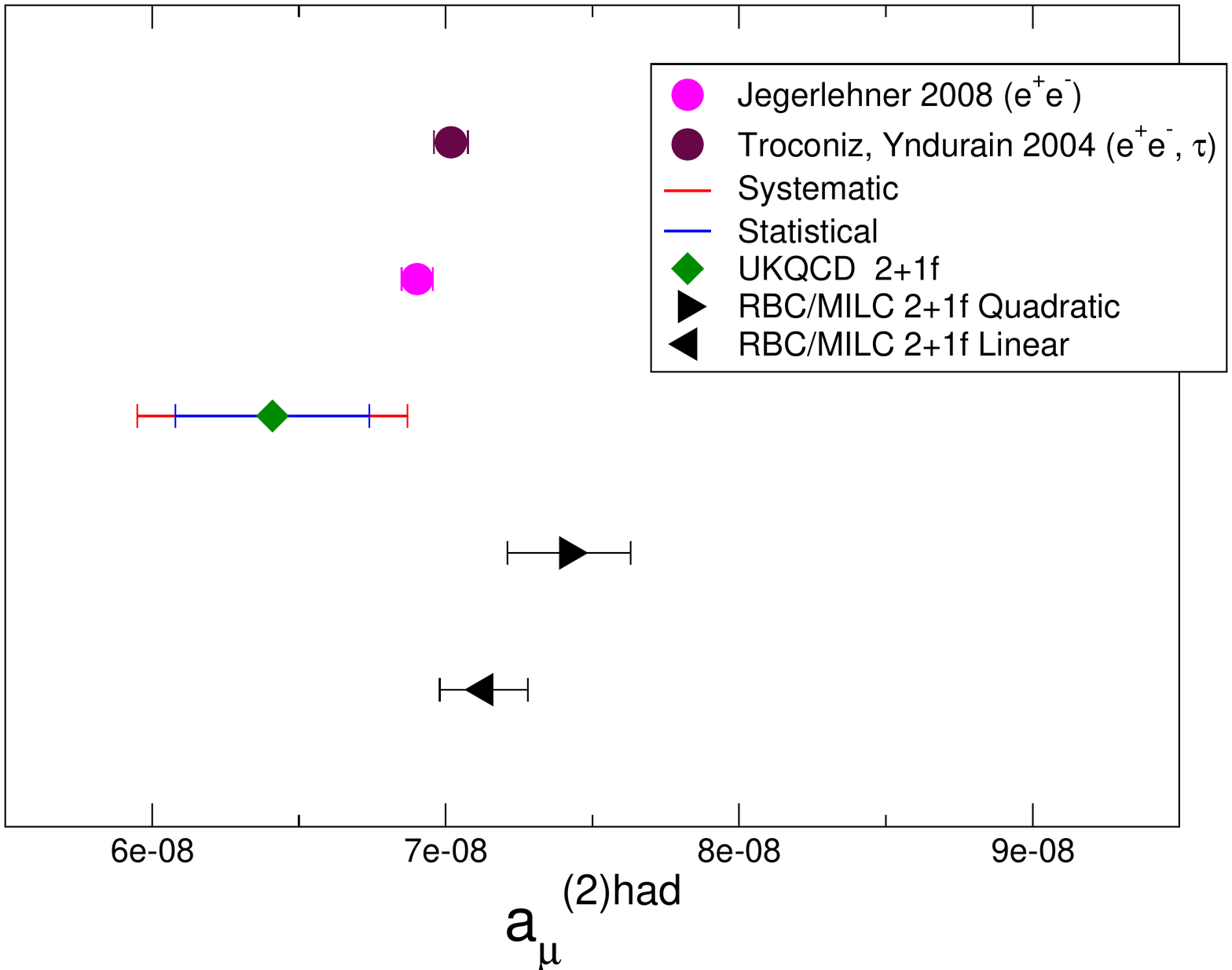}
\label{fig:lineupa}
}
\subfloat[Full value of $a_\mu$.]{
\includegraphics[scale=0.27]{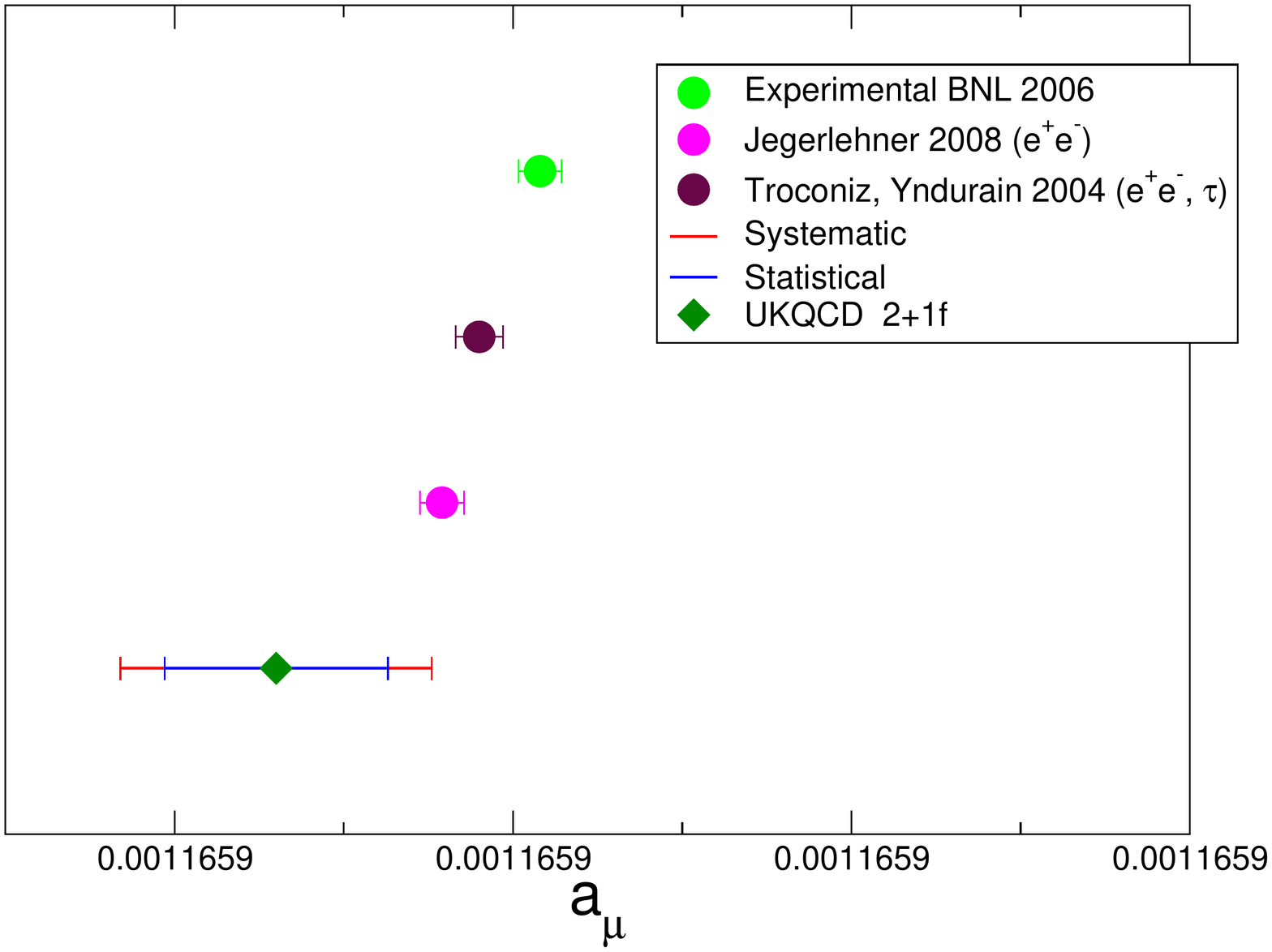}
\label{fig:lineupb}
}
\caption{\small In Fig. \ref{fig:lineupa} we compare recent results for $a_\mu^{(2)had}$.  Our result is compared to previous 2+1f lattice result
\cite{Aubin:2006xv} and results from a dispersive integral over cross-section data from $e^+e^-$ data \cite{Jegerlehner:2008zz} and from $e^+e-$ and $\tau$ data
\cite{Troconiz:2004tr}. In Fig. \ref{fig:lineupb} we compare the full result for $a_\mu$ including the results corresponding to the dispersive results in
Fig. \ref{fig:lineupa} and the current experimental result from \cite{Bennett:2006fi}. The increment between the labelled ticks
on the horizontal axis is $10^{-8}$}\label{rescompplot}
\end{figure}

In Fig. \ref{rescompplot}, our result (\ref{eq:standres}) including a 5\% statistical error arising from the chiral extrapolation (estimated from the
discrepancy between (\ref{eq:standres}) and (\ref{eq:janres}) ) is compared to recent 2+1 flavour lattice results \cite{Aubin:2006xv} 
 along with recent result arising from dispersion integrals over
experimental data from scattering data. Our result is in rough agreement with other results bearing in mind that we have neglected the disconnected
contributions to our correlators, producing a systematic deviation which is bounded to be of the order of 10\%. At this time we are not in a position to
improve on previous theoretical evaluations of $a_\mu^{(2)had}$ and so cannot comment on the scale of the discrepancy between the measured value of $a_\mu$ and
the Standard Model prediction. However it is clear that the next iteration of this calculation with planned improvements is very likely to be in a position to
begin clarifying the situation concerning this discrepancy.

In Table \ref{tab:fv} we attempt a comparison of the value of $F_1$ (defined in (\ref{eq:twovec}) ) resulting from our fit, to the vector decay constant as
measured on each lattice, according to the relation expressed in (\ref{eq:vecpole}). Note, we do not have a result for $f_V$ on the $64\times 24^3$ lattices
at this time, although the ratio of the vector coupling to the vector and tensor currents was studied in \cite{Donnellan:2007xr}. We also make the comparison
suggested by the one-loop correction to
(\ref{eq:vecpole}) as computed in \cite{Aubin:2006xv} whereby the relation $F_1^2\sim\frac{2}{3}f_V^2$ is replaced by $F_1^2\sim\frac{2}{3}f_V^2\times C^2$
where
\begin{equation}
 C^2=1-\frac{6}{(4\pi f_\pi)^2}\left[m_\pi^2 \log\left(\frac{m_\pi^2}{\mu^2}\right)+m_K^2 \log\left(\frac{m_K^2}{\mu^2}\right)\right]
\end{equation}
with $m_\pi$ and $m_K$ the pion and kaon meson masses, $f_\pi$ the pion decay constant, and $\mu$ the chiral scale, taken as 1 GeV. We note that in our fits
we have not included the one-loop contribution from the pseudoscalar sector, and so this comparison can only serve as a rough indication, and we do not
necessarily expect close agreement. We observe that the value of $F_1$ emerging from our fits is on the correct scale when compared to the measured $f_V$,
simply supporting the credibility of our fits. 


\section{Conclusions}

We present a fully dynamical calculation of the leading-order hadronic vacuum polarisation contribution to the anomalous magnetic moment of the muon, using a
2+1 flavour simulation lattice QCD using domain wall fermions. Although we have an expensive fermion discretisation, we improve the accuracy of our result by
convolving an accurate determination of the ground-state vector meson mass with our determination of the lattice hadronic vacuum polarisation in order to
suppress the systematic uncertainty associated with the choice of fit ansatz. Our chiral extrapolation involves lattices at different bare couplings, and thus different lattice spacings, however at this level of precision we do not detect any significant discretisation, or finite volume errors in our result. Our final result we take to be 
\begin{equation}
 a_\mu^{(2)had}=641(33)(32)\times 10^{-10}\label{eq:finalres}
\end{equation}
where the first error is statistical and the second is an estimate of the systematic error arising from the extrapolation to the chiral limit, taken as 5\%, motivated by the variation between the results (\ref{eq:standres}) and (\ref{eq:janres}). Our largest systematic uncertainty arises from the omission of the disconnected contributions and is of the order of 10\% \cite{Juttner:2009yb}.
In order to obtain a more comprehensive and accurate result, we must include the disconnected contributions in our calculation. Furthermore, this being a first
effort at deducing this quantity from our lattices, we have plans to improve it in a number of ways. In addition to the enhancement of our statistics, we would
like to obtain a higher momentum resolution through the use of twisted boundary conditions, and also to explore the use of stochastic sources to further enhance
our signal. With these improvements we would hope to decrease the uncertainty in our result significantly and thus begin to clarify the true discrepancy, if
any, between the actual value of $a_\mu$ and its prediction from the Standard Model.
\begin{table}[!htp]
\centering
\begin{tabular}{c|c|c|c|c|c|c|c}
\hline

$\beta$&$am_u$ &$Q_C^2$ GeV$^2$&$\frac{\chi^2}{n.d.f}$&$a_\mu^{(h)}\times10^{10}$&$aF_1$   &$am_2$  &$aF_2$ \\\hline\hline 
$2.13$ & 0.02  & 4             & 0.38(17)             & 345(16)                  &0.114(4) &1.48(19)&0.31(5)\\\hline
$2.13$ &  0.01 & 3.5           & 0.07(6)              & 430(22)                  &0.110(4) &1.50(23)&0.32(7)\\\hline
$2.13$ & 0.005 & 3.5           & 0.14(5)              & 436(50)                  &0.097(14)&1.16(18)&0.24(3)\\\hline
$2.25$ & 0.008 & 6             & 0.18(11)             & 452(23)                  &0.079(2) &1.14(4) &0.26(1)\\\hline
$2.25$ & 0.006 & 6             & 0.10(6)              & 484(33)                  &0.075(3) &1.07(7) &0.24(2)\\\hline
$2.25$ & 0.004 & 9             & 0.06(3)              & 568(29)                  &0.079(2) &1.23(3) &0.28(6)\\\hline
$1.75$ & 0.0042& 2.5           & 0.16(9)              & 536(36)                  &0.108(20)&1.27(20)&0.26(3)\\\hline
$1.75$ & 0.001 & 2.5           & 0.27(13)             & 646(55)                  &1.06(11) &1.58(61)&0.37(27) \\\hline

\end{tabular}
\caption{\small Results for the hadronic contribution to the muon anomalous magnetic
moment.}
\label{tab:results}
\end{table}

\begin{table}[!htp]
\centering
\begin{tabular}{c|c|c|c|c}
 $\beta$ & $am_u$ & $f_V$ MeV & $\sqrt{\frac{3}{2}}F_1$ MeV  & $\sqrt{\frac{3}{2}}\frac{F_1}{C}$  MeV\\\hline\hline
 2.13    & 0.02   &           & 242(10)                     & 179(7)       \\
 2.13    & 0.01   &           & 234(8)                      & 166(6)      \\
 2.13    & 0.005  &           & 205(30)                     & 144(20)       \\
 2.25    & 0.008  & 178(13)   & 221(6)                      & 155(5)\\
 2.25    & 0.006  & 174(11)   & 211(10)                     & 147(7)\\
 2.25    & 0.004  & 160(26)   & 222(5)                      & 155(4) \\
 1.75    &  0.0042& 140(9)    & 192(27)                     & 129(19)      \\
 1.75    &   0.001& 144(20)   & 179(18)                     & 127(12)
\end{tabular}
\caption{\small Comparison of the vector decay constant as measured on our lattices, to the amplitude of the lowest resonance contribution emerging from our fit to
the lattice vacuum polarisation.}
\label{tab:fv}
\end{table}

\section*{Acknowledgements}
The calculations reported here were performed on the QCDOC computers \cite{Boyle:2005gf,Boyle:2003mj} at Columbia
University, Edinburgh University, and at Brookhaven National Laboratory (BNL), Argonne
Leadership Class Facility (ALCF) BlueGene/P resources at Argonne National Laboratory (ANL), and also the resources of the STFC-funded DiRAC facility. We wish to acknowledge support from STFC grant ST/H008845/1. At BNL,
the QCDOC computers of the RIKEN-BNL Research Center and the USQCD Collaboration were used. The very
large scale capability of the ALCF (supported by the Office of Science of the U.S. Department of
Energy under contract DE-AC02-06CH11357) was critical for carrying out the challenging calculations
reported here. EK is supported by SUPA (The Scottish Universities Physics Alliance). 
JZ is supported by STFC grant ST/F009658/1. This work was supported in part by EU grant 238353 (STRONGnet).
Data used at $\beta=1.75$ is to be presented in an upcoming publication \cite{Kelly:2011up}, and we offer thanks to Chris Kelly for supplying preliminary results on these lattices. 

\bibliographystyle{hunsrt}
\bibliography{refs}

\begin{thebibliography}{10}

\bibitem{Schwinger:1948iu}
Julian~S. Schwinger.
\newblock {On Quantum electrodynamics and the magnetic moment of the electron}.
\newblock {\em Phys. Rev.}, 73:416--417, 1948.

\bibitem{Hanneke:2008tm}
D.~Hanneke, S.~Fogwell, and G.~Gabrielse.
\newblock {New Measurement of the Electron Magnetic Moment and the Fine
  Structure Constant}.
\newblock {\em Phys. Rev. Lett.}, 100:120801, 2008, 0801.1134.

\bibitem{Laporta:2008zz}
S.~Laporta and E.~Remiddi.
\newblock {Status of the QED prediction of the electron (g - 2)}.
\newblock {\em Nucl. Phys. Proc. Suppl.}, 181-182:10--14, 2008.

\bibitem{Bennett:2006fi}
G.~W. Bennett et~al.
\newblock {Final report of the muon E821 anomalous magnetic moment measurement
  at BNL}.
\newblock {\em Phys. Rev.}, D73:072003, 2006, hep-ex/0602035.

\bibitem{Jegerlehner:2009ry}
Fred Jegerlehner and Andreas Nyffeler.
\newblock {The Muon g-2}.
\newblock {\em Phys. Rept.}, 477:1--110, 2009, 0902.3360.

\bibitem{Hayakawa:2005eq}
Masashi Hayakawa, Thomas Blum, Taku Izubuchi, and Norikazu Yamada.
\newblock {Hadronic light-by-light scattering contribution to the muon g-2 from
  lattice QCD: Methodology}.
\newblock {\em PoS}, LAT2005:353, 2006, hep-lat/0509016.

\bibitem{Blum:2009zz}
T.~Blum and S.~Chowdhury.
\newblock {Hadronic contributions to g-2 from the lattice}.
\newblock {\em Nucl. Phys. Proc. Suppl.}, 189:251--256, 2009.

\bibitem{Jegerlehner:2008zz}
F.~Jegerlehner.
\newblock {Muon g - 2 update}.
\newblock {\em Nucl. Phys. Proc. Suppl.}, 181-182:26--31, 2008.

\bibitem{deRafael:1993za}
Eduardo de~Rafael.
\newblock {Hadronic contributions to the muon g-2 and low-energy QCD}.
\newblock {\em Phys. Lett.}, B322:239--246, 1994, hep-ph/9311316.

\bibitem{Blum:2002ii}
T.~Blum.
\newblock {Lattice calculation of the lowest order hadronic contribution to the
  muon anomalous magnetic moment. ((U))}.
\newblock {\em Phys. Rev. Lett.}, 91:052001, 2003, hep-lat/0212018.

\bibitem{Gockeler:2003cw}
{G\"{o}ckeler, M. and others}.
\newblock {Vacuum polarisation and hadronic contribution to muon g-2 from
  lattice QCD}.
\newblock {\em Nucl. Phys.}, B688:135--164, 2004, hep-lat/0312032.

\bibitem{Blum:2003se}
T.~Blum.
\newblock {Lattice calculation of the lowest order hadronic contribution to the
  muon anomalous magnetic moment: An update with Kogut-Susskind fermions}.
\newblock {\em Nucl. Phys. Proc. Suppl.}, 129:904--906, 2004, hep-lat/0310064.

\bibitem{Aubin:2006xv}
C.~Aubin and T.~Blum.
\newblock {Calculating the hadronic vacuum polarization and leading hadronic
  contribution to the muon anomalous magnetic moment with improved staggered
  quarks}.
\newblock {\em Phys. Rev.}, D75:114502, 2007, hep-lat/0608011.

\bibitem{DellaMorte:2010sw}
{Della Morte, Michele and Jager, Benjamin and J\"{u}ttner, Andreas and Wittig,
  Hartmut}.
\newblock {The leading hadronic vacuum polarisation on the lattice}.
\newblock 2010, 1011.5793.

\bibitem{Feng:2011zk}
Xu~Feng, Karl Jansen, Marcus Petschlies, and Dru~B. Renner.
\newblock {Two-flavor QCD correction to lepton magnetic moments at
  leading-order in the electromagnetic coupling}.
\newblock 2011, 1103.4818.

\bibitem{Ohta:2011nv}
Shigemi Ohta.
\newblock {Nucleon structure from 2+1 flavor domain wall QCD at nearly physical
  pion mass}.
\newblock 2011, 1102.0551.

\bibitem{Kelly:2011up}
RBC/UKQCD.
\newblock {Continuum Limit Physics from 2+1 Flavor Domain Wall QCD \Rmnum{2}}.

\bibitem{Allton:2008pn}
C.~Allton et~al.
\newblock {Physical Results from 2+1 Flavor Domain Wall QCD and SU(2) Chiral
  Perturbation Theory}.
\newblock {\em Phys. Rev.}, D78:114509, 2008, 0804.0473.

\bibitem{Aoki:2010dy}
Y.~Aoki et~al.
\newblock {Continuum Limit Physics from 2+1 Flavor Domain Wall QCD}.
\newblock 2010, 1011.0892.

\bibitem{Furman:1994ky}
Vadim Furman and Yigal Shamir.
\newblock {Axial symmetries in lattice QCD with Kaplan fermions}.
\newblock {\em Nucl. Phys.}, B439:54--78, 1995, hep-lat/9405004.

\bibitem{Boyle:2009xi}
Peter~A. Boyle, Luigi Del~Debbio, Jan Wennekers, and James~M. Zanotti.
\newblock {The S Parameter in QCD from Domain Wall Fermions}.
\newblock 2009, 0909.4931.

\bibitem{DellaMorte:2010aq}
{Della Morte, Michele and J\"{u}ttner, Andreas}.
\newblock {Quark disconnected diagrams in chiral perturbation theory}.
\newblock {\em JHEP}, 11:154, 2010, 1009.3783.

\bibitem{Chetyrkin:1996cf}
K.~G. Chetyrkin, Johann~H. Kuhn, and M.~Steinhauser.
\newblock {Three-loop polarization function and O($\alpha(s)^2$) corrections to
  the production of heavy quarks}.
\newblock {\em Nucl. Phys.}, B482:213--240, 1996, hep-ph/9606230.

\bibitem{Arthur:2010ht}
R.~Arthur and P.~A. Boyle.
\newblock {Step Scaling with off-shell renormalisation}.
\newblock {\em Phys. Rev.}, D83:114511, 2011, 1006.0422.

\bibitem{Troconiz:2004tr}
J.~F. de~Troconiz and F.~J. Yndurain.
\newblock {The hadronic contributions to the anomalous magnetic moment of the
  muon}.
\newblock {\em Phys. Rev.}, D71:073008, 2005, hep-ph/0402285.

\bibitem{Donnellan:2007xr}
M.~A. Donnellan et~al.
\newblock {Lattice Results for Vector Meson Couplings and Parton Distribution
  Amplitudes}.
\newblock {\em PoS}, LAT2007:369, 2007, 0710.0869.

\bibitem{Juttner:2009yb}
{J\"{u}ttner, Andreas and Della Morte, Michele}.
\newblock {New ideas for g-2 on the lattice}.
\newblock {\em PoS}, LAT2009:143, 2009, 0910.3755.

\bibitem{Boyle:2005gf}
P.~Boyle et~al.
\newblock {The QCDOC project}.
\newblock {\em Nucl. Phys. Proc. Suppl.}, 140:169--175, 2005.

\bibitem{Boyle:2003mj}
P.~A. Boyle, C.~Jung, and T.~Wettig.
\newblock {The QCDOC supercomputer: Hardware, software, and performance}.
\newblock {\em ECONF}, C0303241:THIT003, 2003, hep-lat/0306023.

\end{thebibliography}
\end{document}